\documentclass[a4paper,11pt]{article}

\pdfoutput=1 
\usepackage{jinstpub} 

\title{\boldmath The laser diode calibration system of
the ICARUS T600 detector at FNAL }

\author[a,1]{M.Bonesini,\note{Corresponding author.}}
\author[a]{R.Benocci,}
\author[a]{R.Bertoni,}
\author[a]{A.Falcone,}
\author[a]{R.Mazza,}
\author[a]{M.Torti,}
\author[b]{A.Menegolli,}
\author[b]{G.L. Raselli,}
\author[b]{M.Rossella} 

\affiliation[a]{Sezione INFN Milano Bicocca, Dipartimento di Fisica 
G. Occhialini, Dipartimento di Scienze dell' Ambiente e della Terra, 
Universit\`a di Milano Bicocca, Milano, Italy}
\affiliation[b]{Sezione INFN Pavia, Dipartimento di Fisica, Pavia, Italy\\}

\emailAdd{maurizio.bonesini@mib.infn.it}

\abstract{ 
The ICARUS T600 LAr TPC is the far detector of the Short Baseline Program 
at FNAL. As it will have to work at shallow depth in the Booster Neutrino Beam,
a large cosmic rays background ($\sim 11$ kHz) will be present.
To reduce it, precise timing information
is needed from the new light detection system, based on 360 
large area photomultipliers. For
 precise time measurements a calibration system based on a fast laser diode 
and a system based on one optical switch, several $1 \times 10$ fused fiber
splitters, ultra-high vacuum  optical feedthroughs  and multimode optical patchcords up to 20 m long, to distribute the
laser pulses to each single PMT, was designed. The time evolution of the
PMTs' gain/timing and possibly their initial calibrations at a time $t_0$ 
will be done by using this system. The expected time resolution of this 
calibration system will be around 100 ps.
 The laboratory tests needed to set up the system are reported.}


\collaboration[c]{ 
on behalf of the ICARUS collaboration}

\begin{document}
\maketitle
\flushbottom
\section{Introduction}
The ICARUS T600 is the largest Liquid Argon Projection Chamber (LAr TPC) in
operation. It took data at LNGS INFN laboratory from 2010 to 2013 on the
CNGS neutrino beam \cite{rubbia}. It was then refurbished at CERN in the
framework of the WA104/NP01 project from 2015 to 2017   and
then moved to FNAL \cite{T600}. 
The improvements introduced during these operations were:
\begin{itemize}
\item{} new cold vessels, with purely passive insulation;
\item{} a renovated cryogenics for the LAr;
\item{} upgrade of the scintillation detector system by using 360 new large
area photomultipliers (PMTs), to provide a time resolution around 1 ns;
\item{} new electronics for the wire chambers' readout.
\end{itemize}
The refurbished ICARUS T600 detector has been installed at
FNAL on the Booster Neutrino Beam (BNB), to search for sterile
neutrinos~\cite{Antonello:2015}.
The new light collection system is based on 360 8" Hamamatsu R5912-MOD 
PMTs ($5\%$ coverage, 15 p.e./MeV), providing sensitivity
down to  $\sim 100$ MeV, good spatial resolution ($\leq 50$ cm), and good
timing resolution ($\sim 1$ ns) \cite{babicz}. The readout is made through 
CAEN V1730 digitizers~\footnote{500 Ms/s, 14 bit, 2 V pp dynamic range, 
in VME standard}.

The time resolution of $\sim 1$ ns will allow to exploit the BNB  bunch structure, for futher rejection of out-of-bunch cosmic
events. To obtain a good  time resolution
 an accurate PMTs' time calibration system 
will be necessary to handle  electronics
time drift due to temperature excursions and other effects.

\section{Layout of the laser calibration system}
\begin{figure}
\centering
\includegraphics[width=0.45\linewidth]{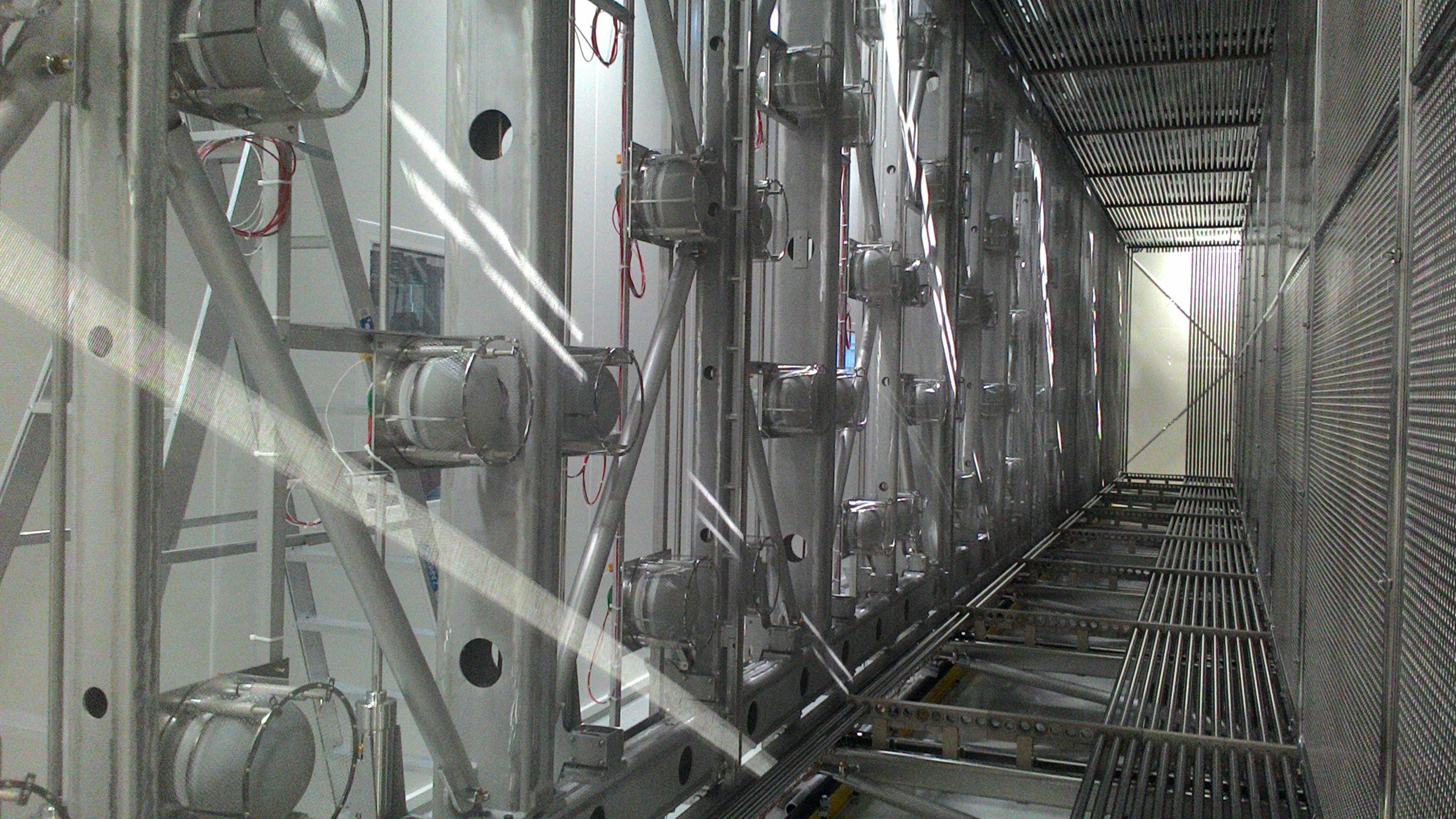}
\includegraphics[width=0.49\linewidth]{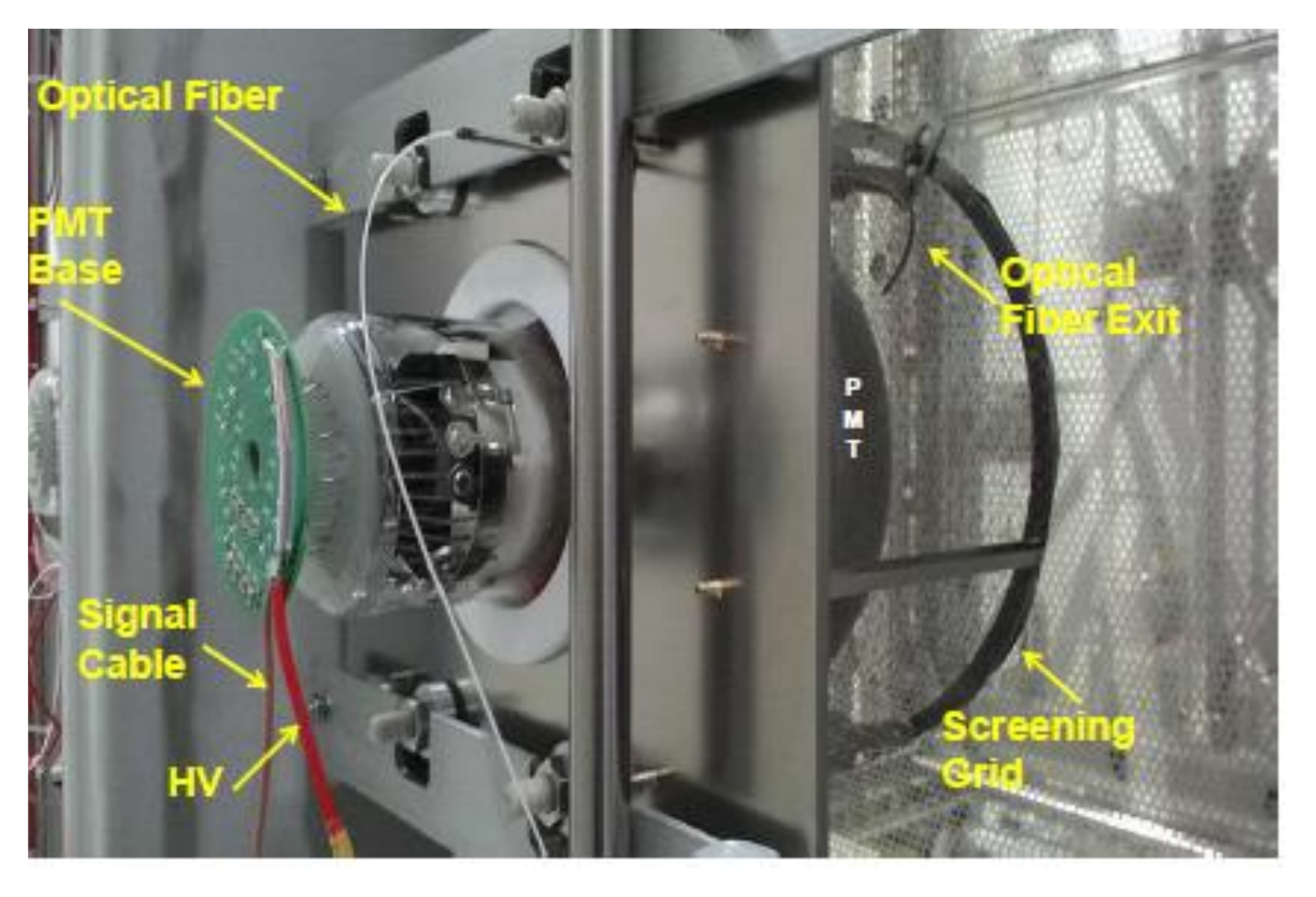}
\caption{Left panel: picture of one of the two sub-modules of the T600
detector, with the 90 large area Hamamatsu PMTs mounted on one side.
Right panel: image of one large area Hamamatsu PMT installed on the TPC frame, 
as seen from the back. Some details are shown.
 }
\label{fig:PMT}
\end{figure}

PMTs' timing calibration  requires the precise
determination of time delays, that may drift for temperature excursions,
  at an initial time $t_0$ and the monitoring of their evolution 
along the data-taking. This may be done with
cosmic rays
 or by delivering a fast calibration pulse
to each individual channel.
\begin{figure}
\centering
\includegraphics[width=0.99\linewidth]{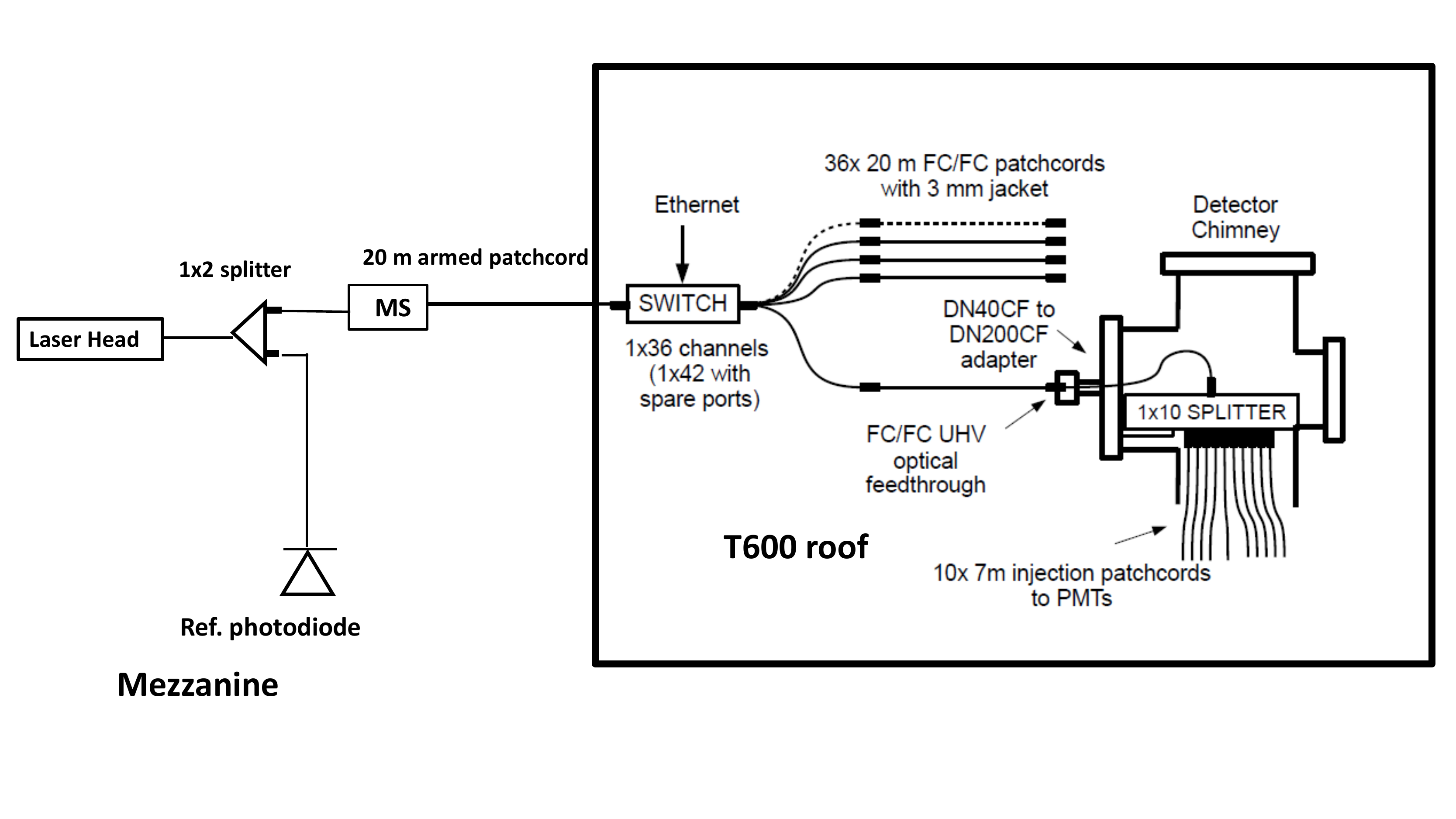}
\caption{Schematic layout of the system proposed for the time calibration
of the 360 PMTs of the ICARUS T600 detector. While the laser diode is 
put in an alcove on a mezzanine of the far detector building, the rest
of the system is positioned on the T600 roof.}
\label{fig:layout}
\end{figure}
The proposed layout for the time calibration of the 360 PMTs of the ICARUS 
T600 detector is outlined  in figure \ref{fig:layout} and is
based on a previous design for the MICE experiment at RAL~\cite{bonesini:2016}.
Fast light pulses from an Hamamatsu PLP10 laser diode~\footnote{ 60 ps FWHM,
100-200 mW peak power, emission at 405 nm}
 are distributed via a 1x36 optical switch
to  20 m long optical patch cables, 
that come 
to optical ultra high vacuum (UHV) 
FC/FC feedthroughs  on CF 40 flanges, mounted on CF40-CF200
adapters.
Inside the T600 tank, 1x10 fused optical splitters are 
attached to each  adapter
to deliver the input laser signal to ten 7 meter long injection fibers that convey the
calibration signal to each PMT, see figure \ref{fig:PMT} for details.
The inner part of the PMTs' calibration system is fully described in
reference \cite{babicz1}.
The laser stability is monitored by a reference photodiode (currently a
Thorlabs DET02AFC)~\footnote{Si photodiode with FC fiber input, 1 ns risetime/falltime, 
1 GHz bandwidth}. 
The main requirement on the calibration system is that the laser pulses
can be delivered to the PMTs' photocathode with a minimal attenuation and
without a deterioration of their original
timing characteristics.
\section{Components characterization}
\begin{figure}
\centering
\includegraphics[width=0.99\linewidth]{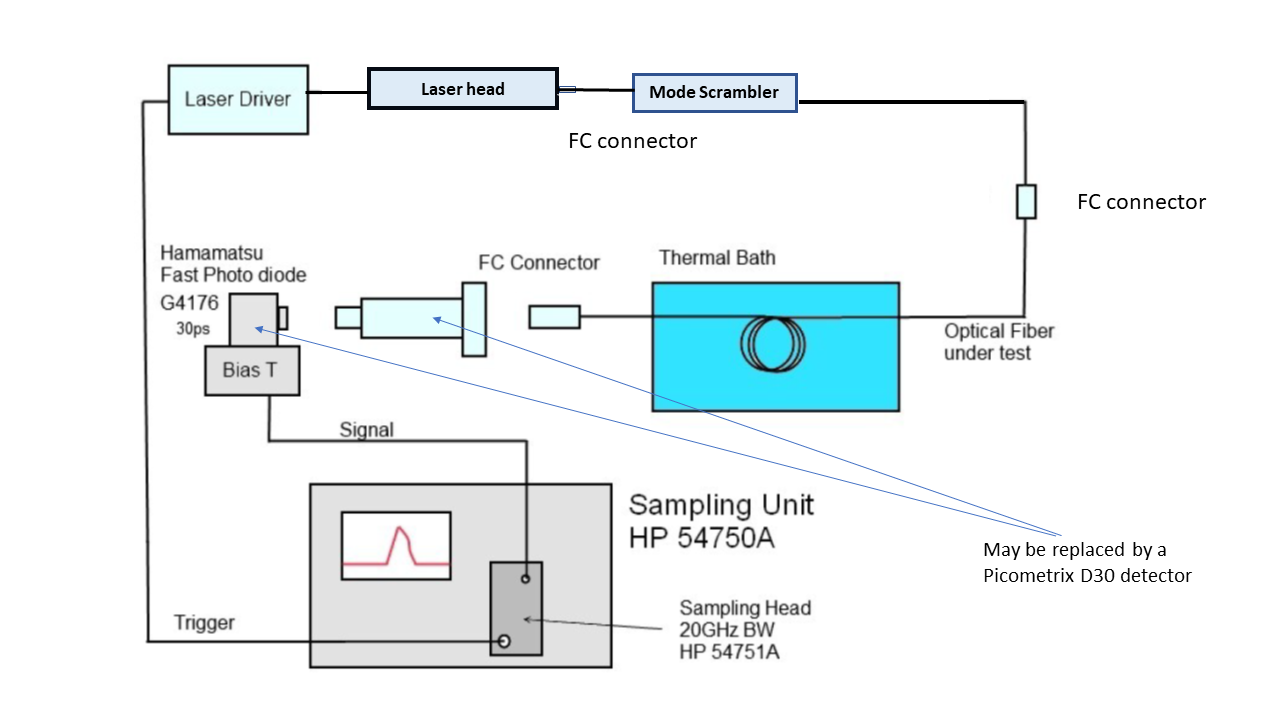}
\caption{Test setup used for characterization of optical components. In part
of the tests the Hamamatsu G4176 photodiode was replaced by a Picometrix
D30 photodetector, with FWHM $\sim 30$ ps. The thermal bath of the Lauda  machine 
was used only for some tests.}
\label{fig:setup}
\end{figure}

The laser pulses
may be sent to one of N output channels by means of an  optical switch and
splitted between M output channels by  suitable fused optical splitters.
Transmission of light between two points is made via optical fiber patches with 
FC/PC  (``fiber channel'') multimode connectors. 
The used optical components (1xN optical switches, 
patch cables, 1xM optical splitters, optical
feedthroughs) were characterized at the used wavelength ($\sim$ 405 nm)
for attenuation and timing properties with a test setup available
at INFN Milano Bicocca (see figure \ref{fig:setup} for details). The main problem is that these components are
easily available at Telecom wavelengths ($\sim 850 - 1300$ nm) but not in the
visible range.
As a consequence, a lot of dedicated tests had to be done. In the test setup, 
the laser light is fed into a ModCon mode scrambler by Arden Photonics~\footnote{insertion loss at 850 nm $\leq $ 3 dB, max power throughput < 10 mW} to ensure
the same distribution of launched modes into the fiber. The optical components
under test (fiber patches or other) 
are then connected in between. The ouput is  analyzed either by a powermeter
\footnote {model OPHIR Nova with PD300 head} 
for transmission tests or fed into a fast detector connected to a sampling 
scope for the analysis of the pulse characteristics, such as risetime/falltime,
FWHM or peak values\footnote{InGaAs G4176 from Hamamatsu or D30 from Picometrix}. 

The main characteristics of the used fast photodetectors are shown in table 
\ref{tab1}. While the G4176-03 photodetector from Hamamatsu is a free space
detector~\footnote{it is packaged in a coaxial metal can, easy to 
connect to an electrical SMA connector}, powered by a Picosecond Pulse 
Lab 5550B bias tee and 
is used together with an Alphalas BBA-3 amplifier, the Picometrix D30 is a 
complete detector with a FC fiber input. 
\begin{table}[ht!]
\centering
\caption{Main characteristics of the used Hamamatsu G4176-03 and Picometrix
D30 detectors. Contributions from bias tee and amplifier are included 
in (*).}
\vspace{.2cm}
\begin{tabular}{|c|c|c|}
\hline
           & Picometrix D30  & Hamamatsu G4176-03 \\ \hline
type       & MSM             & InGaAs MSM \\
FWHM       & 30 ps           & 70 ps (*)  \\
bandwith   & 15 GHz          & $\sim 10$ GHz \\
mount   & fiber input     & free space \\
           & (FC input)      & $0.2 \times 0.2 $ mm$^2$ sensitive area \\
range      & 400-1700 nm     & 450-870 nm \\  
\hline
\end{tabular}
\label{tab1}
\end{table}

The sampling scopes used were either a HP54750A with a HP54751A sampling 
head  or
a Picoscope 9311. Both had a bandwith of 20 GHz, a transition time $\leq 
17.5$ ps and a minimum sensitivity of 1 mV/div. The maximum noise was 
respectively $\leq$ 1 mV and 1.5 mV. 
 \subsection{The laser source}
For a calibration system aiming at a resolution $\sim 0.1$ ns, 
a laser source providing  fast pulses with FWHM around 50 ps is needed. 
Low cost laser sources based on laser diodes have  power
in the range 100 mW - 1 W and therefore injection into the fiber 
delivery system must be optimal. 
Two different laser systems have been used during laboratory tests. 
At first a free space Pilas 040 laser diode from Advanced Laser Systems
was used. Fiber injection was obtained with an Olympus 20x microscope 
objective,
with 0.4 N/A and 1.2 mm working distance, mounted on a x-y-z flexure 
stage with micrometric precision.
Afterwards the system was replaced by an Hamamatsu PLP10 laser diode with
a direct injection system in fiber, via an FC connector mounted on the
laser head. Their main characteristics are summarized in table \ref{tab2}.
\begin{table}[ht!]
\centering
\caption{Main characteristics of the used laser diodes.
}
\vspace{.2cm}
\begin{tabular}{|c|c|c|}
\hline
           & ALS Pilas 040   & Hamamatsu PLP10 \\ \hline
nominal mean power &  1 W            & 120 mW  \\
repetition  rate   &  < 1 MHz        & 2-100 MHz \\
beam head optics & free space & FC fiber input \\
\hline
FWHM      &  $50.7 \pm 0.5$ ps     &  $62.1 \pm 0.1 $ ps \\
risetime  &  $31.2 \pm 0.7$ ps     &  $34.1 \pm 0.1 $ ps \\
pulse height  &  $338 \pm 2$ mV&    $429 \pm 1$ mV \\  
\hline
\end{tabular}
\label{tab2}
\end{table}

\begin{figure}
\centering
\includegraphics[width=0.90\linewidth]{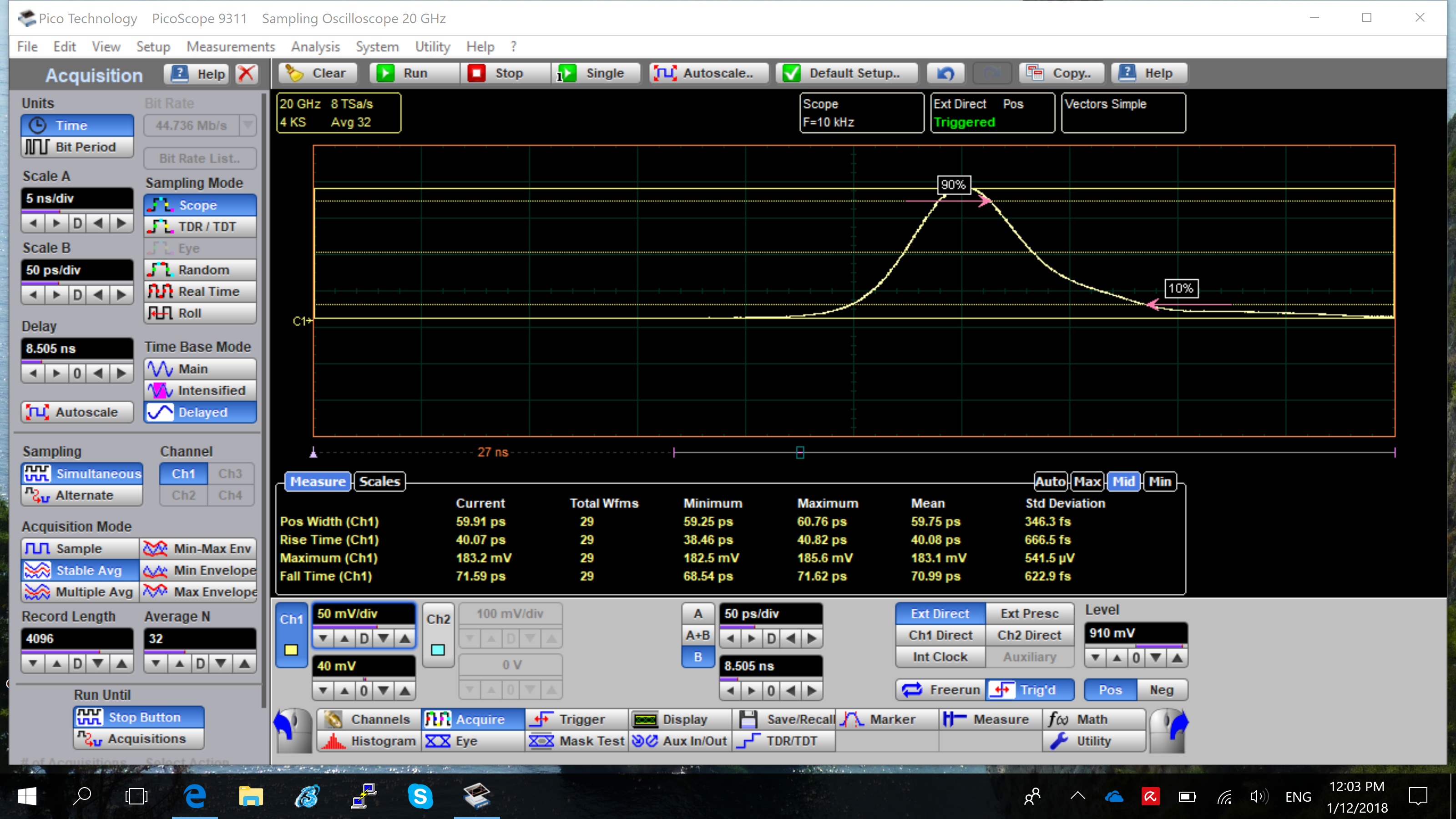}
\caption{Screenshot of laser pulses from Hamamatsu PLP10 laser diode, as seen
on a Picoscope 9311 sampling scope.}
\label{fig-hp}
\end{figure}

FWHM, risetime and max pulse height ($V_{max}$) have been measured 
with an Arden Photonics mode scrambler followed by a Picometrix D30 
fast detector, connected to an HP5475 or a Picoscope 9311 sampling scope
using the setup of figure \ref{fig:setup}. 
Figure \ref{fig-hp} shows a screenshot of laser pulses from the Hamamatsu 
PLP10 laser diode as measured  on a Picoscope 9311 sampling scope, after the 
Picometric D30 fast detector.

The short-time PLP10 laser diode stability was tested over a period of $\sim 3$ hours
of continuos operation and its  power was within $\pm 1\%$ as measured
with a powermeter.
Long-term stability will be monitored instead by a fast photodiode as shown
in figure \ref{fig:layout}.
 \subsection{Optical fibers}
Single mode (SM) optical 
fibers provide negligible dispersion at a given wavelength 
for distances up to some kilometers, but they need very accurate injection 
devices, because of their very limited acceptance angle and core dimensions 
(few times the optimized light wavelength). On the other hand, 
multimode fibers (MM) do
not present injection difficulties but, due to modal dispersion up to 30 ps/m, 
are useless for precise timing over distances larger than 
few meters.
In the visible range a compromise may be found by using MM fibers of 
small core 
diameter, around 50 $\mu$m or less, that behave as limited mode fibers
\cite{tofw}.
Results on the attenuation and timing spread in typical multimode (MM)
50 $\mu$m core
fibers are summarized in table \ref{tab:res1}.
\begin{table}[ht!]
\centering
\caption{Attenuation and timing properties of measured MM fibers with
50 $\mu$m core. S for step index, G for graded index fibers. Dispersion
is measured as the FWHM of the laser pulse.}
\vspace{.2cm}
\begin{tabular}{|c|c|c|c|c|c|}
\hline
fiber type &    N/A & range (nm) &   Att. & $\Delta t$ FWHM  & delay   \\
           &    & &       (dB/m)   & (ps/m)           &  (ns/m) \\ \hline
IRVIS OZ/Optics (G)  &  0.20 & 400-1800 &  0.06   &  0.99            & 5.11 \\
UVVIS OZ/Optics (S)  &  0.22 & 200-900  & 0.12   &  2.21            & 5.05 \\
Corning 50/125 (G)   & 0.20 & 800 -1600 &   0.09    &  1.63            & 6.4  \\
Thorlabs AFS 50/125  (S)     & 0.22 & 400 -2400 & 0.08    &  2.71            & 5.11 \\
Thorlabs SFS 50/125 (S) & 0.22 & 250- 1200& 0.07    &  0.63            &  -  \\
\hline
\end{tabular}
\label{tab:res1}
\end{table}
As optical patches up to 20 m were needed, the choice was the IRVIS OZ/Optics fiber working in
the range 400-1800 nm, with a 0.20 numerical aperture.
\subsection{Optical switches}
An optical switch $1 \times N$ conveys the input signal to one of the N ouput 
channels. Main requirements are a minimal insertion loss (IL) and signal 
equality 
between the different output ports. In addition the fact that the channel to channel cross-talk be minimal is needed.

Different optical switches from PiezoJena, Leoni and Agiltron were tested. 
The choice of a suitable optical switch was the critical point of all
the project. This device is quite common in the Telecom application range
($\lambda \sim 1300$ nm) but it was very difficult to obtain a custom device 
working in the visible range. 

The main characteristics of the optical switches, both provided by the 
producer and measured by us  are
shown in table \ref{tab:switch}.
For the tests the Arden Photonics mode scrambler
was used before the optical switch, followed by a 1m OZ/Optics 
IRVIS patch cord after it.
The laser was operated at 1 MHz and the optical power measured with 
a Thorlabs PM20 powermeter.
All switches, except the one from Agiltron US, had performances far below
the nominal  ones at the reference wavelength 
of 400 nm. 
Their performances, using an Hamamatsu PLP10 laser diode, are reported
in table \ref{tab:switch}. 
\begin{table}[ht!]
\centering
\caption{Main characteristics of tested optical switches.}
\vspace{.2cm}
\begin{tabular}{|c|c|c|c|c|}
\hline
           & PiezoJena  & Leoni MOL   & Leoni MOL  & Agiltron \\
           & F109-05    &  $1 \times 12$ &  $1 \times 36$ & SelfAlign \\
\hline 
fiber type & SFS 50/125  & SIF 50/125 UV & SIF 50/125 UV & IRVIS 50/125  \\
no. ports  &  9                 & 12 &  36  & 46 \\
nominal IL (dB)  &   1.4        & 2.0   & 3.0     & < 0.5 at 400 nm \\
nominal cross-talk (dB)&  -60  & -45    & -45     &  -50  \\
max power (mW)    &    & & & 300 \\
\hline 
IL(dB) at 400 nm &  $4.36 \pm 0.18$   & $1.72 \pm 0.26$ & $4.81 \pm$ 0.90 & 
$0.32 \pm 0.12$  \\
$\frac{FWHM}{FWHM_0}$ & $1.03 \pm 0.01$ & $1.21 \pm 0.02$ & $1.94 \pm 0.09 $ &
$1.03 \pm 0.02$ \\
delay spread (ns) & 0.29 & 1.63 & 0.07 & 0.05 \\
\hline
\end{tabular}
\label{tab:switch}
\end{table}

\begin{figure}
\centering
\includegraphics[width=0.80\linewidth]{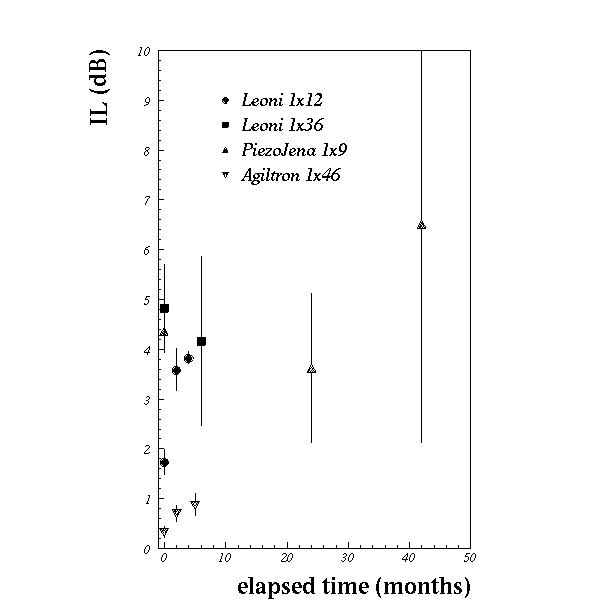}
\caption{ Average insertion loss with RMS spread over the switch pigtails,
as measured at different times from delivery.}
\label{fig:switch1}
\end{figure}

The main problems are related to bad uniformity of the IL between
different pigtails, high IL at 400 nm and unstability of the response vs 
operation time. The evolution of the average insertion loss (IL) as a 
function of the elapsed time from switch delivery is shown in figure 
\ref{fig:switch1}. 
The only optical switch with acceptable performances was
the one provided by Agiltron.  

\subsection{Fused optical splitters}
Extensive tests were done  on $1 \times 10$  fused optical
splitters from Comcore, OZ/Optics and Lightel for uniformity
of output attenuation, delays  and  time spread (FWHM) in the different 
pigtails.  
The total insertion loss (tot. IL) is defined at the power ratio 
between the sum of the output channels and the input one.
The splitting ratio is defined as the power ratio between one channel
output and the total input power. 
Results are reported as averages $\pm$ RMS over the ten channels of a splitter.

A typical fused fiber splitter has a leading pigtail that, due
to the production process, conveys  up to 40-50$\%$ more power
than
the other ones. This effect is amplified at 400 nm, as the production of splitters
is usually made at  higher wavelengths. To reduce this problem an in-line
attenuator is used for this leading channel to equalize the response.

Some test results are reported in table \ref{tab:splitter}.
\begin{table}[ht!]
\centering
\small
\caption{Measured characteristics of typical $1 \times 10$ 
 MM optical splitters. 
Results are reported as averages $\pm$ RMS, to give an idea of the
distribution width over the pigtails of a specimen. In (*) two pigtails
have delays different by more than 10 ns from the others and have been excluded in  the calculation of the delay spread.
}
\vspace{.2cm}
\begin{tabular}{|c|c|c|c|c|c|c|}
\hline
           &  Comcore & Comcore & Lightel & Lightel  & OZ/Optics  & OZ/Optics \\
           & no 1 & no 2 & no 1 & no 2 & no 1 & no 2 \\
\hline
splitting ratio ($\%$) & $ 2.74 \pm 0.69 $ & $ 3.95 \pm 0.40 $ & $5.47 \pm 1.16 $ &
$4.46 \pm 0.90 $ & $ 4.06 \pm 1.34 $& $4.16 \pm 1.23 $ \\
$\frac{FWHM}{FWHM_0}$ & $1.07 \pm 0.03$ & & $1.04 \pm 0.01$ & $1.03 \pm 0.01$ & 
$1.04 \pm 0.01 $ & $1.02 \pm 0.02 $ \\
tot. IL(dB) & 5.66 & 4.03 & 2.62 & 3.50 & 2.62 & 3.81 \\
delay spread (ns) & 0.55 & 0.12 & 0.004 & 0.019  & 0.053 (*) & 0.046 (*) \\
 \hline
\end{tabular}
\label{tab:splitter}
\end{table}

The best solution was obtained with Lightel splitters built for operation at
600 nm with Corning 50/125 fibers.

\subsection{UHV optical feedthrough}

\begin{figure}
\centering
\includegraphics[width=0.45\linewidth]{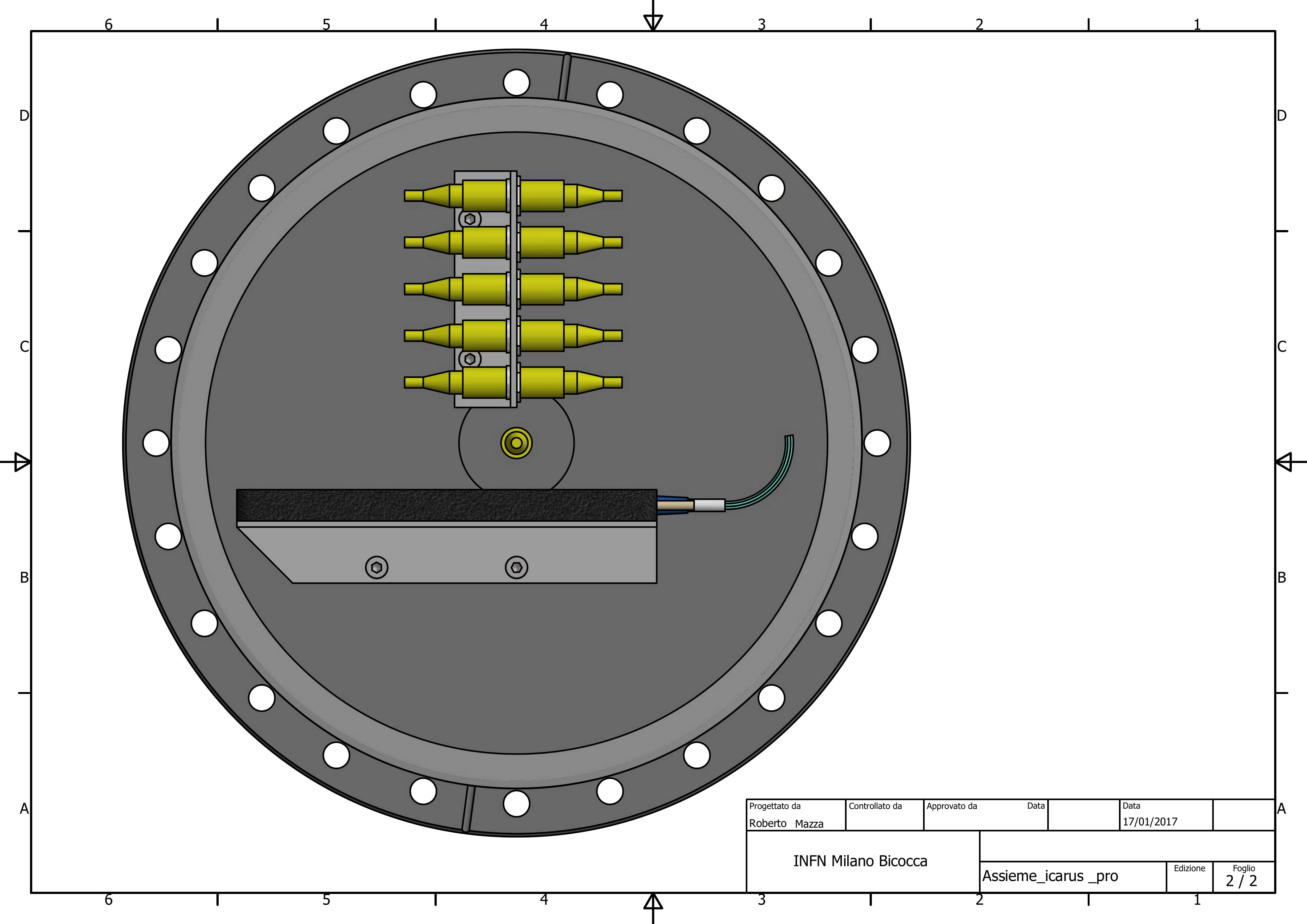}
\includegraphics[width=0.45\linewidth]{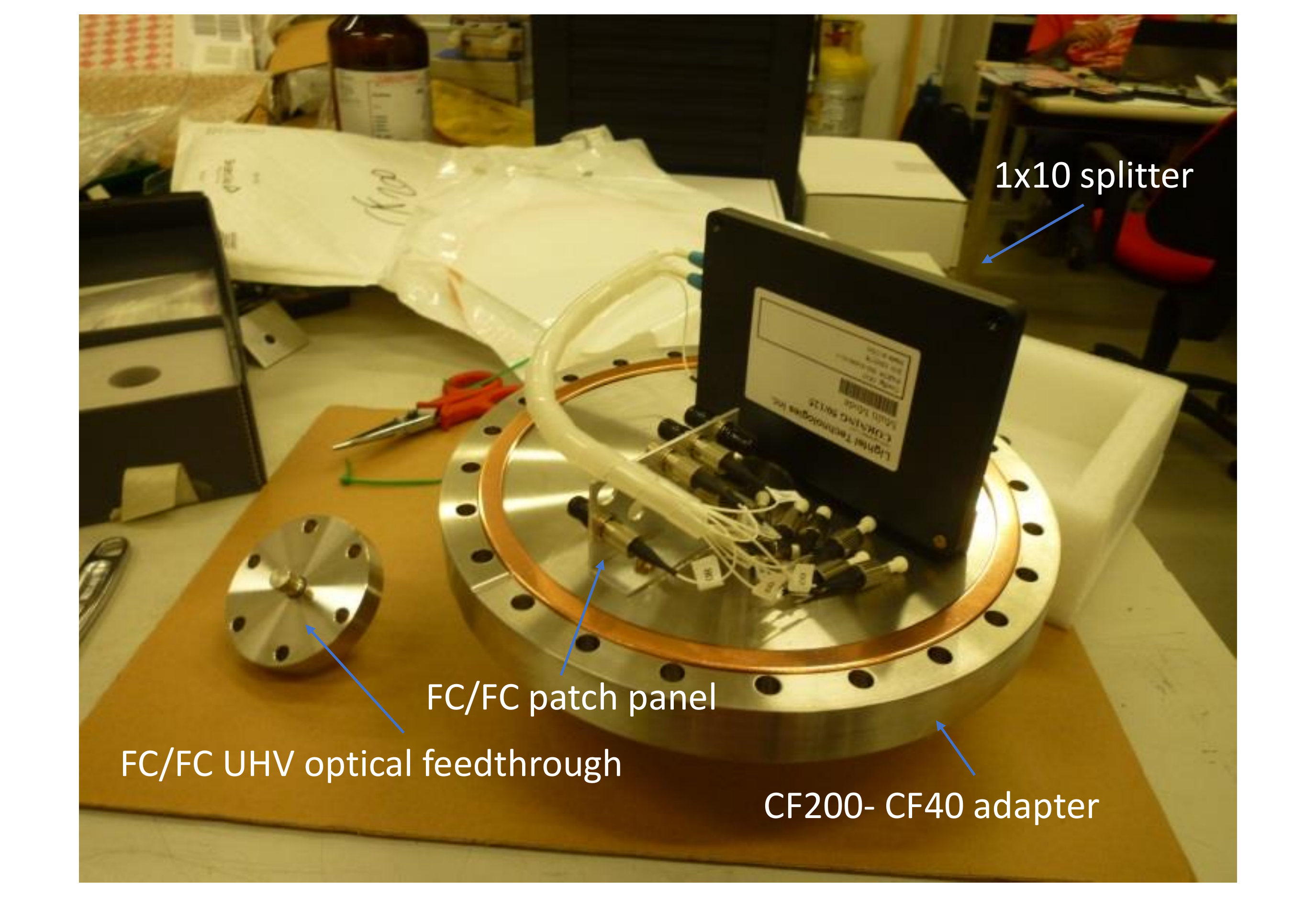}
\caption{Left panel: layout of one optical feedthrough used to convey laser pulses
inside the ICARUS detector as seen from the top. Right panel: picture of the same 
feedthrough with  the CF200-CF40 adapter, 
the 1x10 splitter and  the optical cable patch panel. 
}
\label{fig:feed}
\end{figure}

UHV optical feedthroughs must  convey the
laser pulse inside the ICARUS cryogenic tank. Many different solutions were
tested from glued fiber feedthroughs from OZ/Optics to FC/FC feedthrough realized
with an inner internal fiber. The adopted solution  from VACOM, with a MM
50 $\mu$m core fiber,  had a
measured transmission at 405 nm around 90\%, introduced an additional
delay $\sim 130 $ ps and a negligible  additional time dispersion (FWHM).
Test results are  summarized in table \ref{tab:feed}. 
\begin{table}[ht!]
\centering
\caption{Characteristics of the VACOM optical feedthrough on CF flanges. 
Results are reported as averages $\pm$ RMS.
}
\vspace{.2cm}
\begin{tabular}{|c|c|}
\hline
fiber type & MM $50 \mu m$ core \\
He leak rate & $1 \times 10^{-10} $ mbar l /s \\
\hline

 IL (dB) & $0.38 \pm 0.06$  \\
 $\Delta T$ (ps)& $133.2 \pm 7.4 $ \\ 
 $\Delta (FWHM) (ps) $ &  $0.32 \pm 0.28$      \\
 \hline
\end{tabular}
\label{tab:feed}
\end{table}
To adapt these
flanges to the existing T600 chimneys custom CF40-CF200 adapters from
Lesker Ltd UK are used,
see figure \ref{fig:feed} for details.
On the inner side of the CF40-CF200 adapters, the $1 \times 10$ fused fiber 
splitters were mounted. Their pigtails then go to a FC/FC patch panel, to
which the 7 m long injection patches to the PMTs are connected.
\subsection{Temperature dependence tests}
While the laser, optical switches and 20 m patch cables must work on the roof
of the  ICARUS detector, at room temperature, the 1x10 splitters have to work
inside the T600 chimneys at a slightly lower temperature and the 7m injection
patches at cryogenic temperatures, inside the LAr bath. Tests were done
to assess the dependence from temperature, using a
Lauda thermal machine (precision $0.1$ $^{\circ}$C) in the range 0-50 $^{\circ}$C  and  with
$LN_2$ inside a dewar.
For the chosen fiber IRVIS OZ/Optics
the temperature dependence of the delays is 106 fs/m $^{\circ}$C. For the 7m injection
patch at cryogenic temperature the pulse FWHM increases by 10$\%$.
Other parameters, such as attenuation and delays, are compatible with room temperature ones.
In addition,  the selected splitters $1 \times 10$ were tested in the temperature
range [-10,+15] $^{\circ}$C without no evidence of effects.
\section{Installation of the system at FNAL and preliminary tests}
As all the measurements of the used optical components gave a good transmission
and pulse width spreads well below 50 ps, a resolution of the timing calibration procedure below $\sim $ 100 ps is within reach.

The laser calibration system was deployed on the top of the T600 detector
during 2019. From the laser source to the entrance of the UHV feedtrough,
placed on the chimneys,
a mean attenuation of 4.59 dB with an RMS spread of 0.16 dB was measured.
This value includes contributions from the Mode Scrambler, the 12 m long armed 
patch cable 
going from the laser source  
to the optical switch put on the T600 roof, the optical switch and 
the 20 m long patch cables
going to each flange.
\section{Conclusions}
The layout of the  laser diode calibration system of the ICARUS T600 detector
is described.
The expected performances, such as  a resolution of $\sim 100$ ps,  fit well 
the ICARUS light detection system calibration requirements for
timings at the 1 ns level.

\section*{Acknowledgments}
The help of Mr. S. Banfi and F. Chignoli of INFN Milano Bicocca, Mr. M.C. Prata and Dr. A. de Bari of INFN Pavia,
Ing. G. Manusardi of Fiberlan srl, Dr. M. Wang of Lightel Inc and Dr. T. Wang
of
Agiltron inc 
are gratefully acknowledged.

This work was supported by the EU Horizon 2020 Research and
Innovation Programme under the Marie Sklodowska-Curie Grant
 Agreement No. 822185.

\end{document}